\documentclass[aps,11pt,prd,showpacs,superscriptaddress,preprint]{revtex4}
\usepackage{booktabs,amsmath}
\usepackage{amssymb}
\usepackage{array}
\usepackage{amsthm}
\usepackage{newlfont}
\usepackage{multirow}
\usepackage{graphics}
\usepackage{graphicx}
\usepackage{ae,aecompl}
\usepackage{color}
\usepackage{graphicx}
\usepackage[T1]{fontenc}
\usepackage[latin1]{inputenc}
\usepackage{amsfonts}
\usepackage{bm}
\usepackage[bookmarks=true,bookmarksopen=true,bookmarksnumbered=true,bookmarksopenlevel=3]{hyperref}

\begin{document}

\title{Localized exact solutions of $\mathcal{PT}$ symmetric nonlinear
Schr\"odinger equation with space and time modulated nonlinearities}
\author{\textsc{L. E. Arroyo Meza}}
\author{\textsc{M. B. Hott}}
\author{\textsc{A. de Souza Dutra}}
\affiliation{UNESP Universidade Estadual Paulista - Campus de Guaratinguet\'a - DFQ.\\
Av. Dr. Ariberto Pereira da Cunha, 333 CEP 12516-410, Guaratinguet\'a - SP,
Brazil.}
\author{\textsc{P. Roy}}
\affiliation{Physics and Applied Mathematics Unit, Indian Statistical Institute,
Kolkata-700108, India}
\date{\today }

\begin{abstract}
Using canonical transformations we obtain localized (in space) exact
solutions of the nonlinear Schr\"odinger equation (NLSE) with space and time
modulated nonlinearity and in the presence of an external potential
depending on space and time. In particular we obtain exact solutions of NLSE
in the presence of a number of non Hermitian $\mathcal{PT}$ symmetric
external potentials.
\end{abstract}

\maketitle

\section{Introduction}

Solitons are solitary waves which preserve their shapes during propagation
as well as after collisions. They emerge as solutions of various nonlinear
equations, the nonlinear Schr\"odinger equation (NLSE) (in different
contexts also being called Gross-Pitaevskii equation) being one of them.
Over the years NLSE has found applications in a wide range of fields \cite%
{sulem}. In particular, it plays a crucial role in Bose-Einstein
condensation \cite{BE} and nonlinear optics \cite{nlo,kiv}. It may be noted
that the nonlinearities may be of various types e.g, cubic, quartic, quintic
etc and the couplings or strength management associated with these
nonlinearities may either be a constant or may depend on space, time or
both. Here we shall be concerned with a NLSE with cubic nonlinearity (CNLSE)
and strength management depending on space and time.

In recent years non Hermitian quantum mechanics has been studied in great
detail. The Hamiltonians of many of these systems, in particular the $%
\mathcal{PT}$ symmetric \cite{bender} and the $\eta $ pseudo Hermitian \cite%
{mostafa} ones admit real eigenvalues despite being non Hermitian. The
possibility of realizing $\mathcal{PT}$ symmetric structures in various
fields e.g, optics has been suggested \cite{optic}. It may be mentioned that
lately the study of NLSE with complex $\mathcal{PT}$ symmetric potentials
has drawn a lot of attention because of intrinsic interest as well as for
possible applications. In particular, soliton solutions of the NLSE with
constant nonlinearity and complex potentials have been obtained by many
authors \cite{hu}. In this context it may also be mentioned that schemes of
experimental observation of $\mathcal{PT}$ symmetry has also been reported
\cite{expt}.

Here our objective is to obtain exact localized solutions of NLSE with
space-time modulated nonlinearities \cite{konotop} and in the presence of
complex $\mathcal{PT}$ symmetric potentials. However, in general obtaining
exact solutions of NLSE with space-time modulated nonlinearities are
considerably more difficult in comparison to NLSE with constant
nonlinearities. On the other hand, many solutions of NLSE with constant
nonlinearity are known to exist. So, as a method of solution we shall employ
canonical transformation to map NLSE with space-time modulated nonlinearity
and an external potential onto a NLSE with constant nonlinearity and complex
time independent $\mathcal{PT}$ symmetric potential for which exact
solutions are known. This method was recently employed \cite{lam} to reduce
a NLSE with space and time modulated cubic and quintic nonlinearities into a
stationary NLSE with constants coefficients. Here, once we are interested to
reach a stationary NLSE with a complex $\mathcal{PT}$ symmetric potential,
we resort to a slight modification of the method due to \cite{Jun}.

The organization of the paper is as follows: in section \ref{method} we
discuss the method to transform NLSE with variable nonlinearities to one
with constant nonlinearity and a time independent non Hermitian potential;
in section \ref{examp} we use the formalism described in section \ref{method}
to obtain solutions of NLSE with $\mathcal{PT}$ symmetric external
potentials and finally section \ref{con} is devoted to a conclusion.

\section{Method of solution}

\label{method}

In this section we present the method of solution by focusing on the
non-autonomous CNLSE, namely
\begin{equation}
i\frac{\partial \Psi }{\partial Z}+m\left( X,Z\right) \frac{\partial
^{2}\Psi }{\partial X^{2}}+v\left( X,Z\right) \Psi +g_{3}\left( X,Z\right)
\,|\Psi |^{2}\,\Psi +iw\left( X,Z\right) \Psi =0\,,  \label{eq1}
\end{equation}%
where $\Psi =\Psi \left( X,Z\right) $, $m\left( X,Z\right) $ is the
dispersion parameter, $v\left( X,Z\right) $\ is the trapping potential, $%
g_{3}\left( X,Z\right) $ is the strength management of the cubic
nonlinearity and $w\left( X,Z\right) $ is the gain/loss coefficient. The
specific forms of $m(X,Z),v(X,Z),g_{3}(X,Z)$ and $w(X,Z)$ are taken to be
\begin{equation}
m\left( X,Z\right) =\frac{\zeta \left( Z\right) }{\left( \gamma \left(
Z\right) h\right) ^{2}},  \label{eq2}
\end{equation}%
\begin{eqnarray}
v\left( X,Z\right) &=&\omega _{1}\left( X,Z\right) X^{2}+f_{1}\left(
X,Z\right) X+f_{2}\left( X,Z\right) +\zeta \left( Z\right) V\left[ F\left(
h\right) \right]  \notag \\
&&-\frac{\zeta \left( Z\right) }{h^{2}}\left( \left( \frac{h_{\xi }}{2h}%
\right) ^{2}-\frac{d}{d\xi }\left( \frac{h_{\xi }}{2h}\right) \right) ,
\label{eq3}
\end{eqnarray}%
\begin{equation}
g_{3}\left( X,Z\right) =g\zeta \left( Z\right) h,  \label{eq4}
\end{equation}%
\begin{equation}
w\left( X,Z\right) =f_{3}\left( X,Z\right) X+f_{4}\left( X,Z\right) +\zeta
\left( Z\right) W\left[ F\left( h\right) \right] ,  \label{eq5}
\end{equation}%
where $h$ is an invertible, differentiable and positive function: $h=h\left[
\gamma \left( Z\right) X+\delta \left( Z\right) \right] $, $h_{\xi }=\frac{%
dh(\xi )}{d\xi }|_{\xi =\gamma \left( Z\right) X+\delta \left( Z\right) }$
and $F\left( h\right) $ is a function of $h\left[ \xi \right] $.\ The reason
for choosing the inhomogeneous coefficients and potential in this way is
going to be clarified below.

We now perform the following coordinate transformation and time rescaling
\cite{farinadutra}%
\begin{equation}
X=\frac{\xi }{\overline{\gamma }\left( z\right) }-\frac{\overline{\delta }%
\left( z\right) }{\overline{\gamma }\left( z\right) }\quad ,\quad
Z-Z_{0}=\int_{0}^{z}\frac{dz^{\prime }}{\overline{\zeta }\left( z^{\prime
}\right) }\,,  \label{eq6}
\end{equation}%
with $\overline{\gamma }\left[ z\left( Z\right) \right] =\gamma \left(
Z\right) $, $\overline{\delta }\left[ z\left( Z\right) \right] =\delta
\left( Z\right) $, $\overline{\zeta }\left[ z\left( Z\right) \right] =\zeta
\left( Z\right) $ and, in order to remove the first derivative of $\Psi $
with respect to $\xi $ which arises from the transformation (\ref{eq6}), we
redefine the wavefunction $\Psi $ as%
\begin{equation}
\Psi \left[ X\left( \xi ,z\right) ,Z\left( z\right) \right] =\mathrm{e}%
^{-i\,\alpha \left( \xi ,z\right) }\Phi \left( \xi ,z\right) \,,  \label{eq7}
\end{equation}%
where $\alpha \left( \xi ,z\right) =-a\left( z\right) +\frac{1}{2}%
\int_{0}^{\xi }h^{2}(\xi ^{\prime })\left( \frac{\overline{\gamma }_{z}}{%
\overline{\gamma }}\left( \xi ^{\prime }-\overline{\delta }\right) +%
\overline{\delta }_{z}\right) \,d\xi ^{\prime }$, with $a\left( z\right) $
being an arbitrary function. By substituting the Eqs. (\ref{eq2})-(\ref{eq7}%
) into Eq. (\ref{eq1}) one gets%
\begin{eqnarray}
&&\left. i\,\overline{\zeta }\frac{\partial \Phi }{\partial z}+\frac{%
\overline{\zeta }}{h^{2}}\frac{\partial ^{2}\Phi }{\partial \xi ^{2}}%
+\left\{ \left( \omega _{1}+\frac{\overline{\zeta }\left( h\right) ^{2}}{4}%
\overline{\gamma }_{z}^{2}\right) \left( \frac{\,\xi -\overline{\delta }}{%
\overline{\gamma }}\right) ^{2}+\left( f_{1}+\frac{\overline{\zeta }\left(
h\right) ^{2}}{4}\overline{\gamma }_{z}\overline{\delta }_{z}\right) \left(
\frac{\,\xi -\overline{\delta }}{\overline{\gamma }}\right) \right. \right.
\notag \\
&&\left. \left. +f_{2}+\frac{\overline{\zeta }\left( h\right) ^{2}}{4}%
\overline{\delta }_{z}^{2}+\overline{\zeta }\frac{\partial }{\partial z}%
\left( -a+\frac{1}{2\overline{\zeta }}\int_{0}^{\xi }h^{2}(\xi ^{\prime
})\left( \frac{\overline{\gamma }_{z}}{\overline{\gamma }}\left( \xi
^{\prime }-\overline{\delta }\right) +\overline{\delta }_{z}\right) \,d\xi
^{\prime }\right) +\overline{\zeta }V\left[ F(h)\right] \right. \right.
\notag \\
&&\left. \left. -\frac{\overline{\zeta }}{h^{2}}\left( \left( \frac{h_{\xi }%
}{2h}\right) ^{2}-\frac{d}{d\xi }\left( \frac{h_{\xi }}{2h}\right) \right)
\right\} \Phi +g\,\overline{\zeta }h|\Phi |^{2}\,\Phi +i\left\{ \left( f_{3}-%
\frac{h_{\xi }}{h^{2}}\overline{\zeta }\overline{\gamma }_{z}\right) \left(
\frac{\,\xi -\overline{\delta }}{\overline{\gamma }}\right) \right. \right.
\notag \\
&&\left. \left. +f_{4}-\frac{h_{\xi }\overline{\zeta }}{h^{2}}\overline{%
\delta }_{z}-\frac{\overline{\gamma }_{z}\overline{\zeta }}{2\overline{%
\gamma }}+\overline{\zeta }W\left[ F(h)\right] \right\} \Phi =0\,\right. ,
\label{eq8}
\end{eqnarray}%
where $f_{k}=f_{k}\left( \xi ,z\right) $ ($k=1,2,3,4$).\ From the last
equation one can see why the factors involving $\gamma \left( Z\right) $, $%
\zeta \left( Z\right) $ and $h\left[ \gamma \left( Z\right) X+\delta \left(
Z\right) \right] $ are present in the expressions of $v\left( X,Z\right) $, $%
m\left( X,Z\right) $, $g_{3}\left( X,Z\right) $ and why we have chosen the
specific dependence of $h$, $V$ and $W$ on $\xi =\gamma \left( Z\right)
\,X+\delta \left( Z\right) $. Now, one may choose $\overline{\gamma }\left(
z\right) $, $\overline{\delta }\left( z\right) $, $\overline{\zeta }\left(
z\right) $ and $h(\xi )$\ such that%
\begin{eqnarray}
\omega _{1} &=&-\frac{\overline{\zeta }h^{2}}{4}\overline{\gamma }%
_{z}^{2}\quad ,\quad f_{1}=-\frac{\overline{\zeta }h^{2}}{4}\overline{\gamma
}_{z}\overline{\delta }_{z},  \notag \\
f_{2} &=&-\frac{\overline{\zeta }h^{2}}{4}\overline{\delta }_{z}^{2}-%
\overline{\zeta }\frac{\partial }{\partial z}\left( -\overline{a}+\frac{1}{2%
\overline{\zeta }}\int_{0}^{\xi }h^{2}(\xi ^{\prime })\left( \frac{\overline{%
\gamma }_{z}}{\overline{\gamma }}\left( \xi ^{\prime }-\overline{\delta }%
\right) +\overline{\delta }_{z}\right) \,d\xi ^{\prime }\right) ,  \notag \\
f_{3} &=&\frac{h_{\xi }\overline{\zeta }}{h}\overline{\gamma }_{z}\quad
\mathrm{and}\quad f_{4}=\frac{h_{\xi }}{h}\bar{\zeta}\overline{\delta }_{z}+%
\frac{\overline{\gamma }_{z}\overline{\zeta }}{2\overline{\gamma }}.
\label{eq9}
\end{eqnarray}%
In terms of the original variables ($X,Z$), the functions $\omega
_{1},f_{i},~i=1,2,3,4$ are given by
\begin{eqnarray}
\omega _{1}\left( X,Z\right) &=&-\frac{h^{2}\gamma _{Z}^{2}}{4\zeta \left(
Z\right) },~f_{1}\left( X,Z\right) =-\frac{h^{2}}{2\zeta \left( Z\right) }%
\gamma _{Z}\delta _{Z},  \notag \\
f_{2}\left( X,Z\right) &=&-\frac{\partial }{\partial Z}\left( -a+\frac{%
\gamma }{2\zeta }\int_{0}^{X}h^{2}(\xi ^{\prime })\left( \gamma
_{Z}X^{\prime }+\delta _{Z}\right) \,dX^{\prime }\right) -\frac{h^{2}}{%
4\zeta \left( Z\right) }\delta _{Z}^{2},  \notag \\
f_{3}\left( X,Z\right) &=&\frac{h_{\xi }}{h}\gamma _{Z}\quad \mathrm{and}%
\quad f_{4}=\frac{h_{\xi }}{h}\delta _{z}+\frac{\gamma _{Z}}{2\gamma },
\label{eq10}
\end{eqnarray}%
where $\gamma _{Z}=d\gamma /dZ$, $\delta _{Z}=d\delta /dZ$ \ revealing the
intrinsic connection between $\omega _{1}\left( X,Z\right) $,~ $f_{k}\left(
X,Z\right) $\ ($k=1,2,3,4$) on the functions $\gamma \left( Z\right) $, $%
\delta \left( Z\right) $, $\zeta \left( Z\right) $ and $h\left[ \xi =\gamma
\left( Z\right) X+\delta \left( Z\right) \right] $. Thus, Eq. (\ref{eq8})
takes the form%
\begin{eqnarray}
&&\left. i\frac{\partial \Phi }{\partial z}+\frac{1}{h^{2}}\frac{\partial
^{2}\Phi }{\partial \xi ^{2}}+gh|\Phi |^{2}\,\Phi +iW\left[ F\left( h\right) %
\right] \Phi +\right.  \notag \\
&&\left. +\left\{ V\left[ F\left( h\right) \right] -\frac{1}{h^{2}}\left(
\left( \frac{h_{\xi }}{2h\left[ \xi \right] }\right) ^{2}-\frac{d}{d\xi }%
\left( \frac{h_{\xi }}{2h}\right) \right) \right\} \Phi =0,\right.
\label{eq10b}
\end{eqnarray}%
and the wavefunction (\ref{eq7}) is written as%
\begin{equation}
\Psi \left( X,Z\right) =\mathrm{e}^{-i\,\alpha \left( X,Z\right) }\,\Phi %
\left[ \xi \left( X,Z\right) ,z\left( Z\right) \right] \,,  \label{eq12}
\end{equation}%
where $\alpha \left( X,Z\right) =-a\left( Z\right) +\frac{\gamma \left(
Z\right) }{2\zeta \left( Z\right) }\int_{0}^{X}h^{2}(\xi ^{\prime })\left(
\gamma _{Z}X^{\prime }+\delta _{Z}\right) dX^{\prime }$.

Since we still have a NLSE with inhomogeneous nonlinearity, we are going to
make further transformations in order to arrive at a NLSE with constant
nonlinearity. For that we redefine $\xi $ as a function of another variable $%
x$%
\begin{equation}
\xi -\xi _{0}=\int_{x_{0}}^{x}\frac{dx^{\prime }}{\overline{h}(x^{\prime })},
\label{eq13}
\end{equation}%
where $\overline{h}(x)=h\left[ \xi \left( x\right) \right] $ \noindent and $%
x-x_{0}=F(h)=\int_{\xi _{0}}^{\xi }h(\xi ^{\prime })d\xi ^{\prime }$. In
order to remove the first derivative of $\Phi $ with respect to $x$ which
arises from the transformation (\ref{eq13}), we redefine the wavefunction $%
\Phi $ as%
\begin{equation}
\Phi \left( \xi \left( x\right) ,z\right) =\frac{\psi \left( x,z\right) }{%
\sqrt{\overline{h}(x)}}\,.  \label{eq14}
\end{equation}%
By substituting Eqs. (\ref{eq13}) and (\ref{eq14}) into Eq. (\ref{eq10b})
one gets%
\begin{equation}
i\frac{\partial \psi }{\partial z}+\frac{\partial ^{2}\psi }{\partial x^{2}}%
+[V\left( x\right) +iW\left( x\right) ]\psi +g|\psi |^{2}\psi =0,
\label{eq15}
\end{equation}%
where $\psi =\psi \left( x,z\right) $, $V\left( x\right) =V\left[ F(h)\right]
$ and $W\left( x\right) =W\left[ F(h)\right] $ . Note that the variable $z$
can be formally identified with time $t$ and Eq. (\ref{eq15}) can be
interpreted as a NLSE with complex potential $V(x)+iW(x)$. We are now going
to consider cases for which the potential $V(x)+iW(x)$ is invariant under
the $\mathcal{PT}$ transformation, that is $x\rightarrow -x$, $z\rightarrow
-z$ and $i\rightarrow -i$.

By returning to the original space-time coordinates ($X,Z$), the
wavefunction can be obtained from Eqs. (\ref{eq14}) and (\ref{eq12}), that is%
\begin{equation}
\Psi \left( X,Z\right) =\frac{\mathrm{e}^{-i\,\alpha \left( X,Z\right) }}{%
\sqrt{h\left[ \xi \left( X,Z\right) \right] }}\psi \left( x,Z\right) .
\label{eq16}
\end{equation}

Thus, we have shown, by means of point canonical transformations, how the
non-autonomous NLSE with cubic nonlinearity management, Eq. (\ref{eq1}), can
be mapped onto a NLSE with cubic homogeneous nonlinearity, Eq. (\ref{eq15}).

\subsection{Particular case $h\left[ \protect\xi \right] =1$.}

In this case the non-autonomous CNLSE (\ref{eq1}) is written as%
\begin{equation}
i\frac{\partial \Psi }{\partial Z}+m\left( Z\right) \frac{\partial ^{2}\Psi
}{\partial X^{2}}+v\left( X,Z\right) \Psi +g_{3}\left( Z\right) \,|\Psi
|^{2}\,\Psi +iw\left( X,Z\right) \Psi =0\,.  \label{eq16a}
\end{equation}%
The coefficients and the potential reduce to%
\begin{equation}
m\left( Z\right) =\frac{\zeta \left( Z\right) }{\gamma ^{2}\left( Z\right) },
\label{eq16c}
\end{equation}%
\begin{equation}
g_{3}\left( Z\right) =g\zeta \left( Z\right) ,  \label{eq16e}
\end{equation}%
\begin{equation}
w\left( X,Z\right) =\frac{\gamma _{Z}}{2\gamma }+\zeta \left( Z\right)
W\left( \xi \right) ,  \label{eq16f}
\end{equation}

\begin{eqnarray}
v\left( X,Z\right) &=&-\left( \frac{\gamma _{Z}^{2}}{\zeta }+\frac{\partial
}{\partial Z}\left( \frac{\gamma \gamma _{Z}}{\zeta }\right) \right) \frac{%
X^{2}}{4}-\left( \frac{\gamma _{Z}\delta _{Z}}{\zeta }+\frac{\partial }{%
\partial Z}\left( \frac{\gamma \delta _{Z}}{\zeta }\right) \right) \frac{X}{2%
}+  \notag \\
&&+\frac{da}{dZ}-\frac{\delta _{Z}^{2}}{4\zeta }+\zeta \left( Z\right)
V\left( \xi \right) ,  \label{eq16d}
\end{eqnarray}%
where $\xi =\gamma \left( Z\right) X+\delta \left( Z\right) $. From Eq. (\ref%
{eq13}) we deduce that $\xi =x$. Thus, by means of the Eq. (\ref{eq6}) and
the wavefunction (\ref{eq16}), which can be rewritten as
\begin{equation}
\Psi \left( X,Z\right) =\psi \left( x,Z\right) ~\exp \left\{ -i\left( \frac{%
\gamma \gamma _{Z}}{4\zeta }X^{2}+\frac{\gamma \delta _{Z}}{2\zeta }%
X-a\left( Z\right) \right) \right\} ,  \label{eq16b}
\end{equation}%
we can map the non-autonomous CNLSE with cubic nonlinearity management (\ref%
{eq16a}) onto a NLSE with cubic homogenous nonlinearity (\ref{eq15}).

\section{Examples}

\label{examp}

In this section we present some examples. We deal with some specific cases
for which the exact solutions. Explicitly, we take $h\left[ \xi \right] =%
\mathrm{e}^{\xi ^{2}/b^{2}}$, $\zeta \left( Z\right) =\gamma ^{2}$\ , $%
\delta \left( Z\right) =a\left( Z\right) =0$ and $\gamma \left( Z\right) =%
\frac{\gamma _{0}}{\varepsilon +\cos \left( \nu \,Z\right) }$ with $%
\left\vert \varepsilon \right\vert >1$.

These functions are related to the dispersion parameter $m\left( X,Z\right) $
and cubic nonlinearity $g_{3}\left( X,Z\right) $ by Eqs. (\ref{eq2}) and (%
\ref{eq4}), that is
\begin{equation}
m\left( X,Z\right) =\mathrm{e}^{-\frac{2\gamma _{0}^{2}}{b^{2}\left(
\varepsilon +\cos \left( \nu Z\right) \right) ^{2}}X^{2}},~g_{3}\left(
X,Z\right) =\frac{\gamma _{0}^{2}\,g}{\left( \varepsilon +\cos \left( \nu
Z\right) \right) ^{2}}\mathrm{e}^{\frac{\gamma _{0}^{2}}{b^{2}\left(
\varepsilon +\cos \left( \nu Z\right) \right) ^{2}}X^{2}}.  \label{eq17a}
\end{equation}%

It is important to remark that with
this choice and from Eqs. (\ref{eq6}) and (\ref{eq13}), we obtain the
relationship between the original variables $(X,Z)$ and the variables
$(x,z)$, namely
\begin{equation*}
x=\frac{b\sqrt{\pi }}{2}\text{Erfi}\left[ \frac{\gamma _{0}}{b\left(
\varepsilon +\cos \left( \nu Z\right) \right) }X\right] ,
\end{equation*}\begin{equation*}
z=\gamma _{0}^{2}\left( \frac{2\varepsilon }{\nu \left( \varepsilon
^{2}-1\right) ^{3/2}}\arctan \left[ \frac{\varepsilon -1}{\sqrt{\varepsilon
^{2}-1}}\tan \left( \frac{\nu }{2}Z\right) \right] -\frac{\sin \left( \nu
Z\right) }{\nu \left( \varepsilon ^{2}-1\right) \left( \varepsilon +\cos
\left( \nu Z\right) \right) }\right) ,
\end{equation*}where Erfi is the imaginary error function \cite{abramowitz}. Then, we
notice that $(X,Z)\rightarrow(-X,-Z)$ implies into $(x,z)\rightarrow(-x,-z)$, such that we can establish the $\mathcal{PT}$ symmetry
in the examples we are going to treat below.

In the particular case $h\left[ \xi \right] =1$, and with the same functions $%
\gamma \left( Z\right) =\frac{\gamma _{0}}{\varepsilon +\cos \left( \nu
\,Z\right) }$, $\zeta \left( Z\right) =\gamma ^{2}$\ , $\delta \left(
Z\right) =a\left( Z\right) =0$,
the
relationship between the original variables $(X,Z)$ and the variables
$(x,z)$ are
\begin{equation*}
x=\frac{\gamma _{0}}{
\varepsilon +\cos \left( \nu Z\right)}X ,
\end{equation*}\begin{equation*}
z=\gamma _{0}^{2}\left( \frac{2\varepsilon }{\nu \left( \varepsilon
^{2}-1\right) ^{3/2}}\arctan \left[ \frac{\varepsilon -1}{\sqrt{\varepsilon
^{2}-1}}\tan \left( \frac{\nu }{2}Z\right) \right] -\frac{\sin \left( \nu
Z\right) }{\nu \left( \varepsilon ^{2}-1\right) \left( \varepsilon +\cos
\left( \nu Z\right) \right) }\right) .
\end{equation*}
Thus, when $(X,Z)\rightarrow(-X,-Z)$ then $(x,z)\rightarrow(-x,-z)$ too.
Moreover, from Eqs. (\ref{eq16c}) and (\ref{eq16e}) we get%
\begin{equation}
m\left( Z\right) =1,~g_{3}\left( Z\right) =\frac{g\,\gamma _{0}^{2}}{\left(
\varepsilon +\cos \left( \nu Z\right) \right) ^{2}},  \label{eq17b}
\end{equation}%
such that, from (\ref{eq17a}) and (\ref{eq17b}), one has: $m\left( X,Z\right) =m\left(
-X,-Z\right) $ and $g_{3}\left( X,Z\right) =g_{3}\left( -X,-Z\right) .$

\subsection{Example 1}

As a first example, we consider a $\mathcal{PT}$ symmetric potential of
Scarf II type \cite{ahmed}%
\begin{equation}
V(x)=V_{0}\,\mathrm{sech}^{2}\left( x\right) \quad ,\quad W(x)=W_{0}\,%
\mathrm{sech}\left( x\right) \tanh (x)\,,  \label{eq18}
\end{equation}%
where the coupling constants satisfy the condition
\begin{equation}
W_{0}\leq V_{0}+1/4.
\end{equation}%
Then Eq. (\ref{eq15}), with $g=1$, admits a solution of the form%
\begin{equation}
\psi (x,z)=\sqrt{2-V_{0}+\left( \frac{W_{0}}{3}\right) ^{2}}\mathrm{sech}%
\left( x\right) \mathrm{e}^{i\left( z+\frac{W_{0}}{3}\arctan \left[ \sinh
\left( x\right) \right] \right) },  \label{eq19}
\end{equation}%
corresponding to zero boundary condition at $x\rightarrow \pm \infty $.
Substituting the Scarf II type potential, Eq. (\ref{eq18}), in Eqs. (\ref%
{eq3}) and (\ref{eq5}) we obtain the trapping potential $v(X,Z)$%
\begin{eqnarray}
v\left( X,Z\right) &=&-\frac{\left( 4\gamma _{0}^{4}\mathrm{e}^{-\frac{%
2\gamma _{0}^{2}}{b^{2}\left( \epsilon +\cos \left( \nu Z\right) \right) ^{2}%
}X^{2}}+3b^{4}\nu ^{2}\sin ^{2}\left( \nu Z\right) \left( \epsilon +\cos
\left( \nu Z\right) \right) ^{2}\mathrm{e}^{\frac{2\gamma _{0}^{2}}{%
b^{2}\left( \epsilon +\cos \left( \nu Z\right) \right) ^{2}}X^{2}}\right) }{%
4b^{4}\left( \epsilon +\cos \left( \nu Z\right) \right) ^{4}}X^{2}+  \notag
\\
&&+\frac{\gamma _{0}^{2}\mathrm{e}^{-\frac{2\gamma _{0}^{2}}{b^{2}\left(
\epsilon +\cos \left( \nu Z\right) \right) ^{2}}X^{2}}}{b^{2}\left( \epsilon
+\cos \left( \nu Z\right) \right) ^{2}}-\frac{b^{2}\nu ^{2}\left( \epsilon
\cos \left( \nu Z\right) +\cos \left( 2\nu Z\right) \right) }{8\gamma
_{0}^{2}}\left( \mathrm{e}^{\frac{2\gamma _{0}^{2}}{b^{2}\left( \epsilon
+\cos \left( \nu Z\right) \right) ^{2}}X^{2}}-1\right) +  \notag \\
&&+\frac{V_{0}\gamma _{0}^{2}}{\left( \epsilon +\cos \left( \nu Z\right)
\right) ^{2}}\mathrm{\mathrm{sech}}^{2}\left[ \frac{\sqrt{\pi }b}{2}\mathrm{%
Erfi}\left[ \frac{\gamma _{0}}{b\left( \epsilon +\cos \left( \nu Z\right)
\right) }X\right] \right] ,  \label{eq19a}
\end{eqnarray}%
and the gain/loss coefficient $w\left( X,Z\right) $

\begin{eqnarray}
w\left( X,Z\right) &=&\frac{2\gamma _{0}^{2}\nu \sin \left( \nu Z\right) }{%
b^{2}\left( \epsilon +\cos \left( \nu Z\right) \right) ^{3}}X^{2}+\frac{\nu
\sin \left( \nu Z\right) }{2\left( \epsilon +\cos \left( \nu Z\right)
\right) }+\frac{W_{0}\gamma _{0}^{2}}{\left( \epsilon +\cos \left( \nu
Z\right) \right) ^{2}}\times  \notag \\
&&\times \mathrm{\mathrm{sech}}\left[ \frac{\sqrt{\pi }b}{2}\mathrm{Erfi}%
\left[ \frac{\gamma _{0}}{b\left( \epsilon +\cos \left( \nu Z\right) \right)
}X\right] \right] \tanh \left[ \frac{\sqrt{\pi }b}{2}\mathrm{Erfi}\left[
\frac{\gamma _{0}}{b\left( \epsilon +\cos \left( \nu Z\right) \right) }X%
\right] \right] .  \label{eq20}
\end{eqnarray}

\noindent The latter expressions satisfy the following equalities $v\left(
X,Z\right) =v\left( -X,-Z\right) $ and $w\left( X,Z\right) =-w\left(
-X,-Z\right) $, i. e. the trapping potential and gain/loss coefficient are
even and odd respectively, with regard to sign reversal of $X$ and $Z$.
Therefore, we can say that $v\left( X,Z\right) +\,i\,w\left( X,Z\right) $
works as a complex $\mathcal{PT}$ \ symmetric potential and, due to the
specific choices of $h\left[ \xi \right] $, $\gamma \left( Z\right) $, $%
\zeta \left( Z\right) $ and $\delta \left( Z\right) $, the non-autonomous
NLSE in Eq. (\ref{eq1}) is invariant under \textquotedblleft
time\textquotedblright\ ($Z\rightarrow -Z$, $i\rightarrow -i$) and space ($%
X\rightarrow -X$) reversals.

\begin{figure}[h]
\centering
\includegraphics[width=7.0cm]{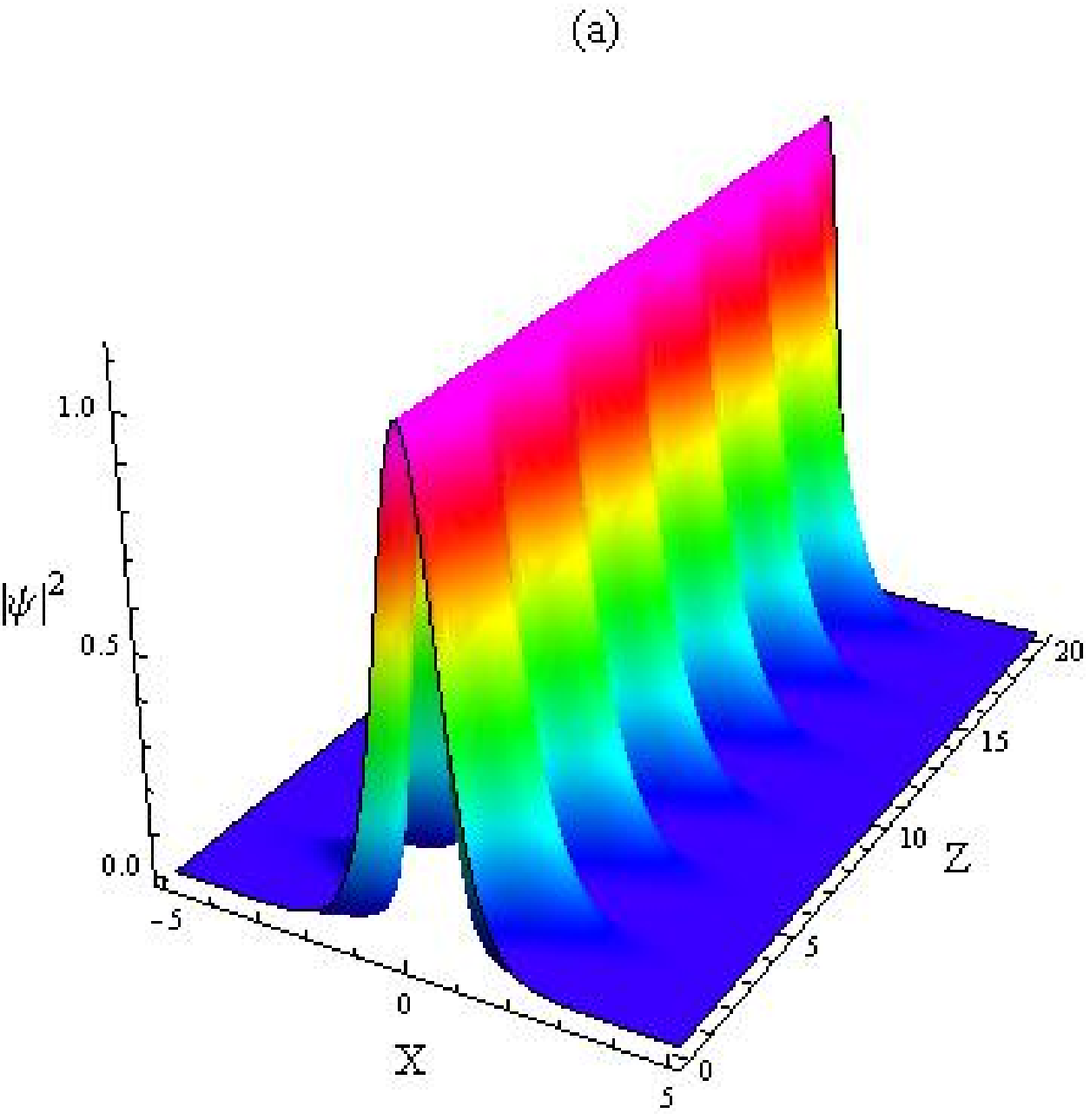}\qquad\qquad %
\includegraphics[width=7.0cm]{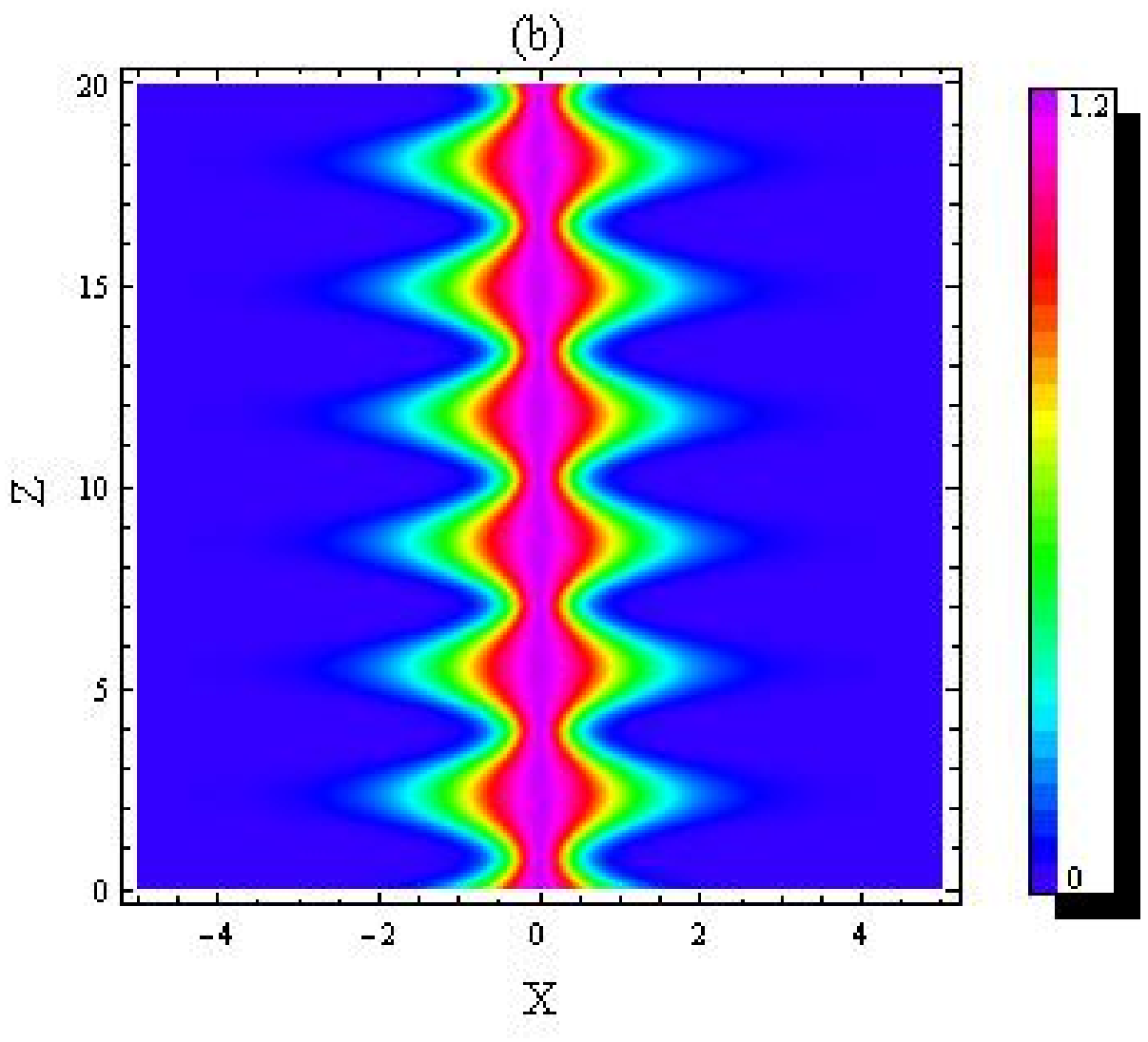}
\caption{(a) $\left\vert \Psi \right\vert ^{2}$ of the bright soliton. (b)
The periodic evolution of bright soliton is better seen from its density
plot. Eq. (\protect\ref{eq21}) with $b=10,$ $\protect\nu =2,$ $\protect%
\epsilon =2,$ $\protect\gamma _{0}=2.5,$ $V_{0}=1$ and $W_{0}=1$. }
\label{fig2}
\end{figure}

The wavefunction $\Psi \left( X,Z\right) $, which is solution of Eq. (\ref%
{eq1}), is obtained by substituting Eq. (\ref{eq19}) into Eq. (\ref{eq16})%
\begin{eqnarray}
\Psi \left( X,Z\right) &=&\sqrt{2-V_{0}+\left( \frac{W_{0}}{3}\right) ^{2}}%
\mathrm{\mathrm{sech}}\left[ \frac{\sqrt{\pi }b}{2}\mathrm{Erfi}\left[ \frac{%
\gamma _{0}}{b\left( \epsilon +\cos \left( \nu Z\right) \right) }X\right] %
\right] \times  \notag \\
&&\times \mathrm{e}^{-\frac{\gamma _{0}^{2}}{2b^{2}\left( \epsilon +\cos
\left( \nu Z\right) \right) ^{2}}X^{2}}\mathrm{e}^{i\varphi \left(
X,Z\right) },  \label{eq21}
\end{eqnarray}%
where
\begin{eqnarray}
\varphi \left( X,Z\right) &=&\int_{0}^{Z}\frac{\gamma _{0}^{2}}{\left(
\epsilon +\cos \left( \nu Z^{\prime }\right) \right) ^{2}}dZ^{\prime }+\frac{%
W_{0}}{3}\arctan \left[ \mathrm{\sinh }\left[ \frac{\sqrt{\pi }b}{2}\mathrm{%
Erfi}\left[ \frac{\gamma _{0}}{b\left( \epsilon +\cos \left( \nu Z\right)
\right) }X\right] \right] \right] +  \notag \\
&&+\frac{b^{2}\nu \left( \epsilon +\cos \left( \nu Z\right) \right) \sin
\left( \nu Z\right) }{8\gamma _{0}^{2}}\left( \mathrm{e}^{\frac{2\gamma
_{0}^{2}}{b^{2}\left( \epsilon +\cos \left( \nu Z\right) \right) ^{2}}%
X^{2}}-1\right) .  \label{eq21a}
\end{eqnarray}%
The plot of the $\left\vert \Psi \left( X,Z\right) \right\vert ^{2}$ is
shown in Fig. (\ref{fig2}).

\begin{figure}[h]
\centering\includegraphics[width=7.0cm]{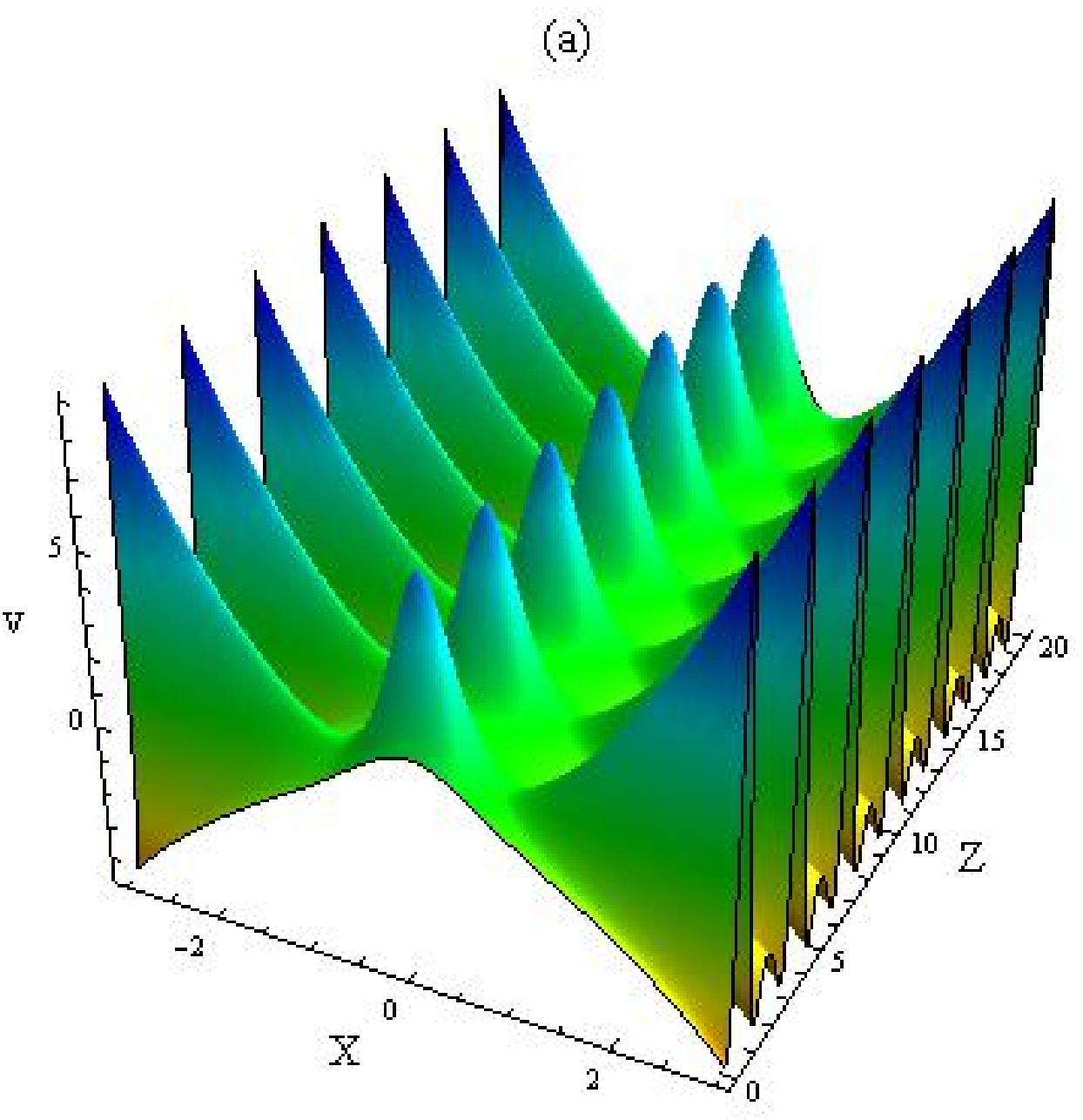}\qquad\qquad %
\includegraphics[width=7.0cm]{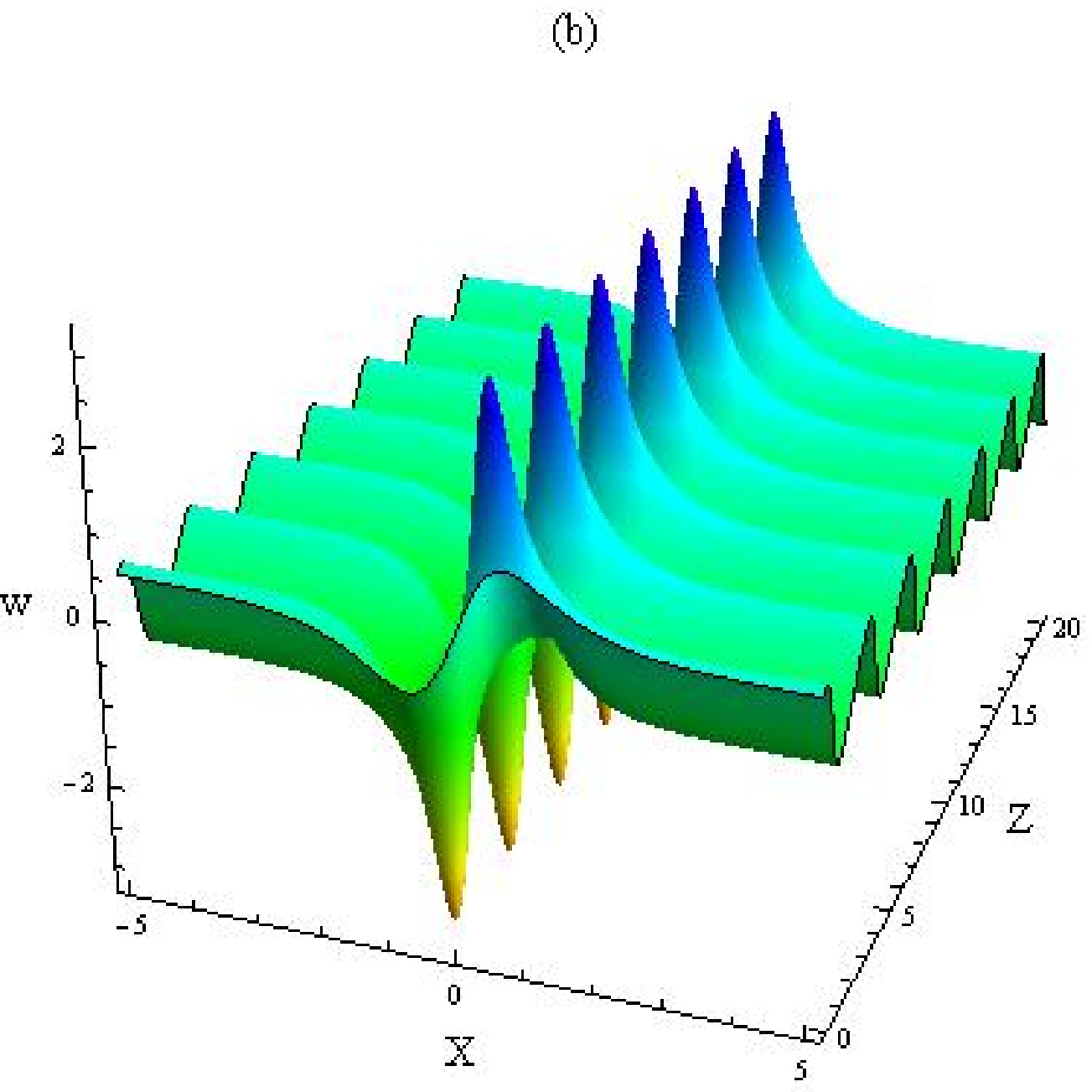}
\caption{Periodic evolution of :(a) the even trapping potential $v\left(
X,Z\right) $ and \ (b) the odd gain/loss coefficient $w\left( X,Z\right) $.
Eq. (\protect\ref{eq21b}) with $\protect\nu =2,$ $V_{0}=1,$ $W_{0}=1$ and $%
\protect\gamma _{0}=2.5$. }
\label{fig1}
\end{figure}

In the case when $h\left[ \xi \right] =1,m(Z)=1$, the trapping potential $%
v(X,Z)$ and the gain/loss coefficient $w\left( X,Z\right) $ become

\begin{eqnarray}
v\left( X,Z\right) &=&\frac{\nu ^{2}(\cos \left( 2\nu Z\right) -3-2\epsilon
\cos \left( \nu Z\right) )}{8\left( \epsilon +\cos \left( \nu Z\right)
\right) ^{2}}X^{2}+\frac{V_{0}\gamma _{0}^{2}}{\left( \epsilon +\cos \left(
\nu Z\right) \right) ^{2}}\mathrm{sech}^{2}\left[ \frac{\gamma _{0}~X}{%
\epsilon +\cos \left( \nu Z\right) }\right] ,  \notag \\
w\left( X,Z\right) &=&\frac{\nu \sin \left( \nu Z\right) }{2\left( \epsilon
+\cos \left( \nu Z\right) \right) }+\frac{W_{0}\gamma _{0}^{2}}{\left(
\epsilon +\cos \left( \nu Z\right) \right) ^{2}}\mathrm{sech}\left[ \frac{%
\gamma _{0}~X}{\epsilon +\cos \left( \nu Z\right) }\right] \tanh \left[
\frac{\gamma _{0}~X}{\epsilon +\cos \left( \nu Z\right) }\right] \,.
\label{eq21b}
\end{eqnarray}%
These last expressions also satisfy $v\left( X,Z\right) =v\left(
-X,-Z\right) $ and $w\left( X,Z\right) =-w\left( -X,-Z\right) $ i. e. the
trapping potential and gain/loss coefficient are even and odd respectively,
with regard to $X$ and $Z$. Thus the potential $v\left( X,Z\right)
+i\,w\left( X,Z\right) $ is $\mathcal{PT}$ symmetric. The plots of the $%
v\left( X,Z\right) $ and $w\left( X,Z\right) $ are shown in Fig. (\ref{fig1}%
).

\noindent The wavefunction $\Psi \left( X,Z\right) $, which is solution of
Eq. (\ref{eq16a}), is obtained from Eq. (\ref{eq16b})%
\begin{equation}
\Psi \left( X,Z\right) =\sqrt{2-V_{0}+\left( \frac{W_{0}}{3}\right) ^{2}}%
\mathrm{\mathrm{sech}}\left[ \frac{\gamma _{0}}{\epsilon +\cos \left( \nu
Z\right) }X\right] \mathrm{e}^{i\varphi \left( X,Z\right) },  \label{eq21c}
\end{equation}%
where%
\begin{equation*}
\varphi \left( X,Z\right) =-\frac{\nu \sin \left( \nu Z\right) }{4\left(
\epsilon +\cos \left( \nu Z\right) \right) }X^{2}+\int_{0}^{Z}\frac{\gamma
_{0}^{2}}{\left( \epsilon +\cos \left( \nu Z^{\prime }\right) \right) ^{2}}%
dZ^{\prime }+\frac{W_{0}}{3}\arctan \left[ \mathrm{\sinh }\left[ \frac{%
\gamma _{0}}{\epsilon +\cos \left( \nu Z\right) }X\right] \right] .
\end{equation*}%
The evolution and behavior of $\left\vert \Psi \left( X,Z\right) \right\vert
^{2}$, in this case, is similar to that shown in Fig. (\ref{fig2}).

\subsection{Example 2}

\begin{figure}[h]
\centering
\includegraphics[width=7.0cm]{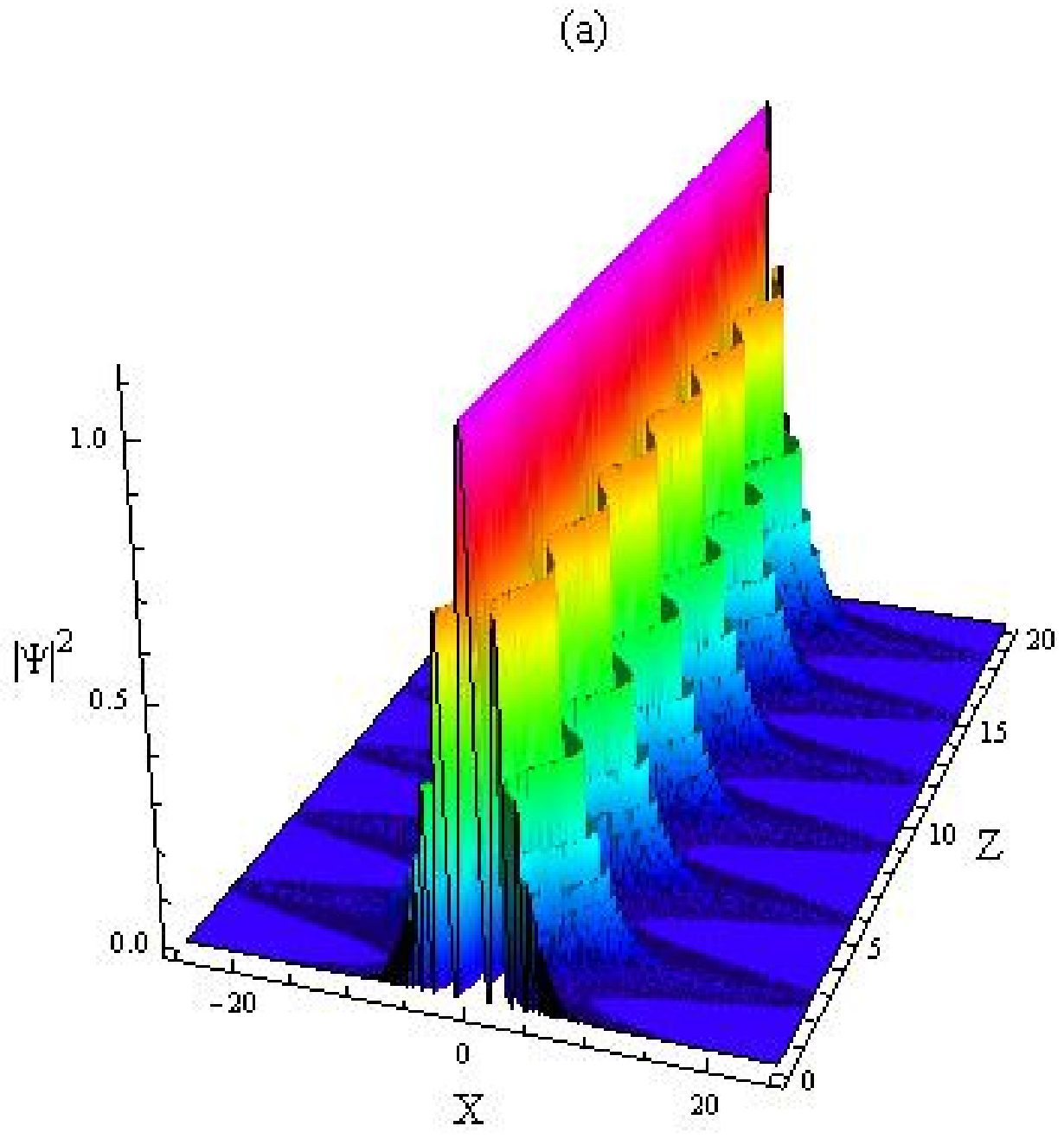}\qquad\qquad %
\includegraphics[width=7.0cm]{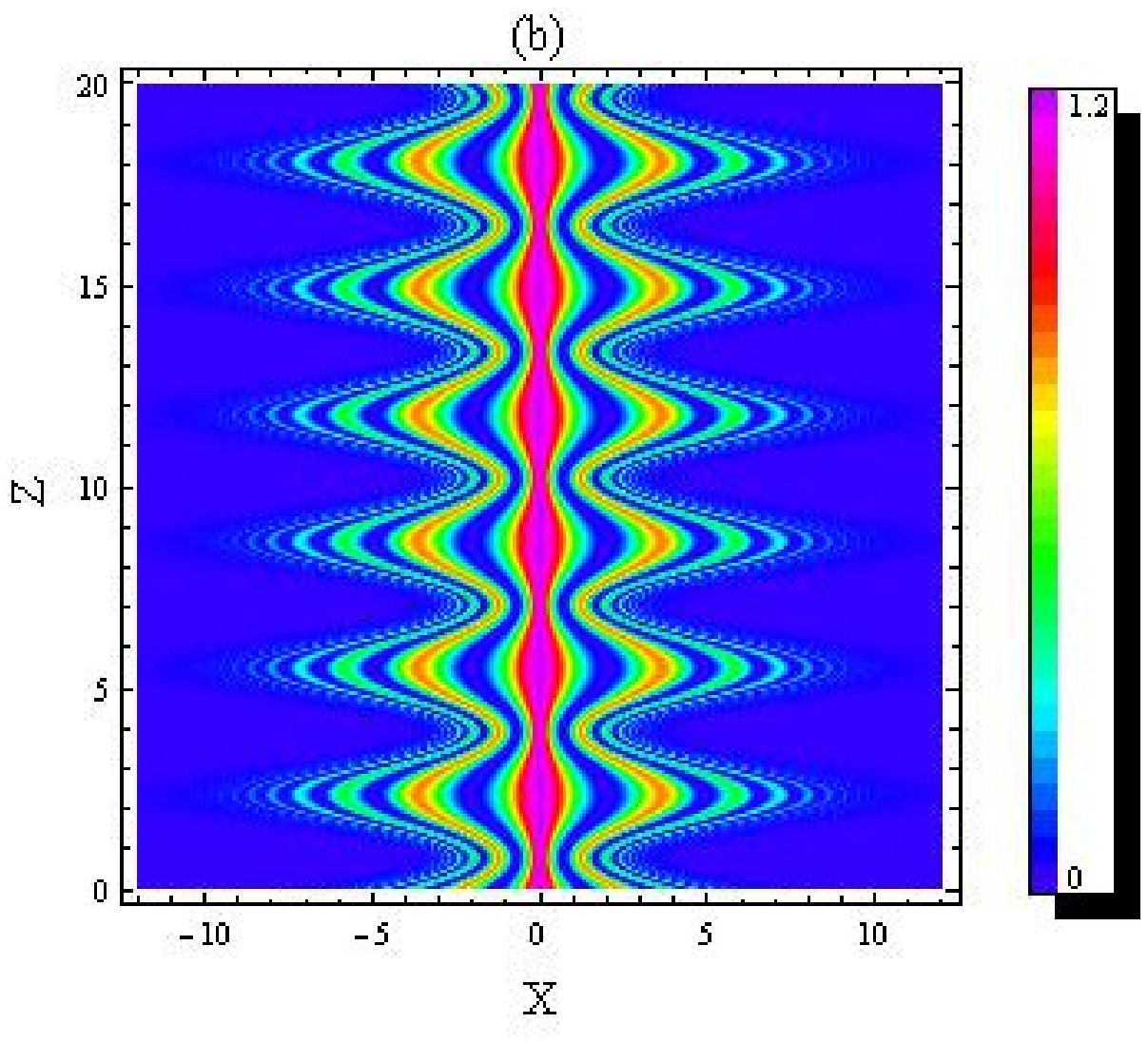}
\caption{(a) Evolution and behavior of $\left\vert \Psi \right\vert ^{2}$ of
the soliton. (b) The periodic evolution of the soliton is better seen from
its density plot. Eq. (\protect\ref{eq24}) with $b=5,$ $\protect\nu =2,$ $%
\protect\epsilon =2;$ $\protect\gamma _{0}=2.5,$ $V_{0}=0.5,$ $W_{0}=0.5$
and $k=0.4$. }
\label{fig4}
\end{figure}

Now we consider a $\mathcal{PT}$ symmetric periodic potential of the form
\cite{muss}%
\begin{equation}
V(x)=V_{0}\mathrm{sn}^{2}\left( x,k\right) +W_{0}^{2}k^{2}\mathrm{sn}%
^{4}\left( x,k\right) \,,\,W(x)=W_{0}\mathrm{sn}\left( x,k\right) \left( 4%
\mathrm{dn}^{2}\left( x,k\right) -1+k^{2}\right) \,.  \label{eq22}
\end{equation}%
The solution $\psi (x,z)$ of Eq. (\ref{eq15}), for such Jacobi periodic
potentials and $g=1$, is of the form%
\begin{equation}
\psi (x,z)=\sqrt{V_{0}+W_{0}^{2}k^{2}+W_{0}^{2}+2k^{2}}\,\mathrm{cn}\left(
x,k\right) \mathrm{e}^{i\left[ \left( V_{0}+W_{0}^{2}k^{2}+2k^{2}-1\right)
z+W_{0}\mathrm{sn}\left( x,k\right) \right] },  \label{eq23}
\end{equation}%
and it exists in the branch $V_{0}>-\left(
W_{0}^{2}k^{2}+W_{0}^{2}+2k^{2}\right) .$ By substituting the Jacobian
periodic potentials, Eq. (\ref{eq22}), in Eqs. (\ref{eq3}) and (\ref{eq5})
we obtain the trapping potential $v(X,Z)$ and the gain/loss coefficient $%
w\left( X,Z\right) $%
\begin{eqnarray}
v\left( X,Z\right) &=&-\frac{\left( 4\gamma _{0}^{4}\mathrm{e}^{-\frac{%
2\gamma _{0}^{2}}{b^{2}\left( \epsilon +\cos \left( \nu Z\right) \right) ^{2}%
}X^{2}}+3b^{4}\nu ^{2}\sin ^{2}\left( \nu Z\right) \left( \epsilon +\cos
\left( \nu Z\right) \right) ^{2}\mathrm{e}^{\frac{2\gamma _{0}^{2}}{%
b^{2}\left( \epsilon +\cos \left( \nu Z\right) \right) ^{2}}X^{2}}\right) }{%
4b^{4}\left( \epsilon +\cos \left( \nu Z\right) \right) ^{4}}X^{2}+  \notag
\\
&&+\frac{\gamma _{0}^{2}\mathrm{e}^{-\frac{2\gamma _{0}^{2}}{b^{2}\left(
\epsilon +\cos \left( \nu Z\right) \right) ^{2}}X^{2}}}{b^{2}\left( \epsilon
+\cos \left( \nu Z\right) \right) ^{2}}-\frac{b^{2}\nu ^{2}\left( \epsilon
\cos \left( \nu Z\right) +\cos \left( 2\nu Z\right) \right) }{8\gamma
_{0}^{2}}\left( \mathrm{e}^{\frac{2\gamma _{0}^{2}}{b^{2}\left( \epsilon
+\cos \left( \nu Z\right) \right) ^{2}}X^{2}}-1\right) +  \notag \\
&&+\frac{\gamma _{0}^{2}}{\left( \epsilon +\cos \left( \nu Z\right) \right)
^{2}}\left( V_{0}+W_{0}^{2}k^{2}\mathrm{sn}^{2}\left[ \frac{\sqrt{\pi }b}{2}%
\mathrm{Erfi}\left[ \frac{\gamma _{0}}{b\left( \epsilon +\cos \left( \nu
Z\right) \right) }X\right] ,k\right] \right) \times  \notag \\
&&\times \mathrm{sn}^{2}\left[ \frac{\sqrt{\pi }b}{2}\mathrm{Erfi}\left[
\frac{\gamma _{0}}{b\left( \epsilon +\cos \left( \nu Z\right) \right) }X%
\right] ,k\right] ,  \notag \\
w\left( X,Z\right) &=&\frac{2\gamma _{0}^{2}\nu \sin \left( \nu Z\right) }{%
b^{2}\left( \epsilon +\cos \left( \nu Z\right) \right) ^{3}}X^{2}+\frac{\nu
\sin \left( \nu Z\right) }{2\left( \epsilon +\cos \left( \nu Z\right)
\right) }+\frac{W_{0}\gamma _{0}^{2}}{\left( \epsilon +\cos \left( \nu
Z\right) \right) ^{2}}\mathrm{sn}\left[ \frac{\sqrt{\pi }b}{2}\mathrm{Erfi}%
\left[ \frac{\gamma _{0}}{b\left( \epsilon +\cos \left( \nu Z\right) \right)
}X\right] ,k\right] \times  \notag \\
&&\times \left( 4\mathrm{dn}^{2}\left[ \frac{\sqrt{\pi }b}{2}\mathrm{Erfi}%
\left[ \frac{\gamma _{0}}{b\left( \epsilon +\cos \left( \nu Z\right) \right)
}X\right] ,k\right] +k^{2}-1\right) .  \label{eq23a}
\end{eqnarray}%
The wavefunction $\Psi \left( X,Z\right) $, which is solution of Eq. (\ref%
{eq1}) is obtained by substituting Eq. (\ref{eq23}) into the Eq. (\ref{eq16})%
\begin{eqnarray}
\left\vert \Psi \left( X,Z\right) \right\vert &=&\sqrt{%
V_{0}+W_{0}^{2}k^{2}+W_{0}^{2}+2k^{2}}\,\mathrm{e}^{-\frac{\gamma _{0}^{2}}{%
2b^{2}\left( \epsilon +\cos \left( \nu Z\right) \right) ^{2}}X^{2}}\times
\notag \\
&&\times \left\vert \mathrm{cn}\left( \frac{\sqrt{\pi }b}{2}\mathrm{Erfi}%
\left[ \frac{\gamma _{0}}{b\left( \epsilon +\cos \left( \nu Z\right) \right)
}X\right] ,k\right) \right\vert .  \label{eq24}
\end{eqnarray}%
The plot of the $\left\vert \Psi \left( X,Z\right) \right\vert ^{2}$ is
shown in Fig. (\ref{fig4}).

\begin{figure}[h]
\centering\includegraphics[width=7.0cm]{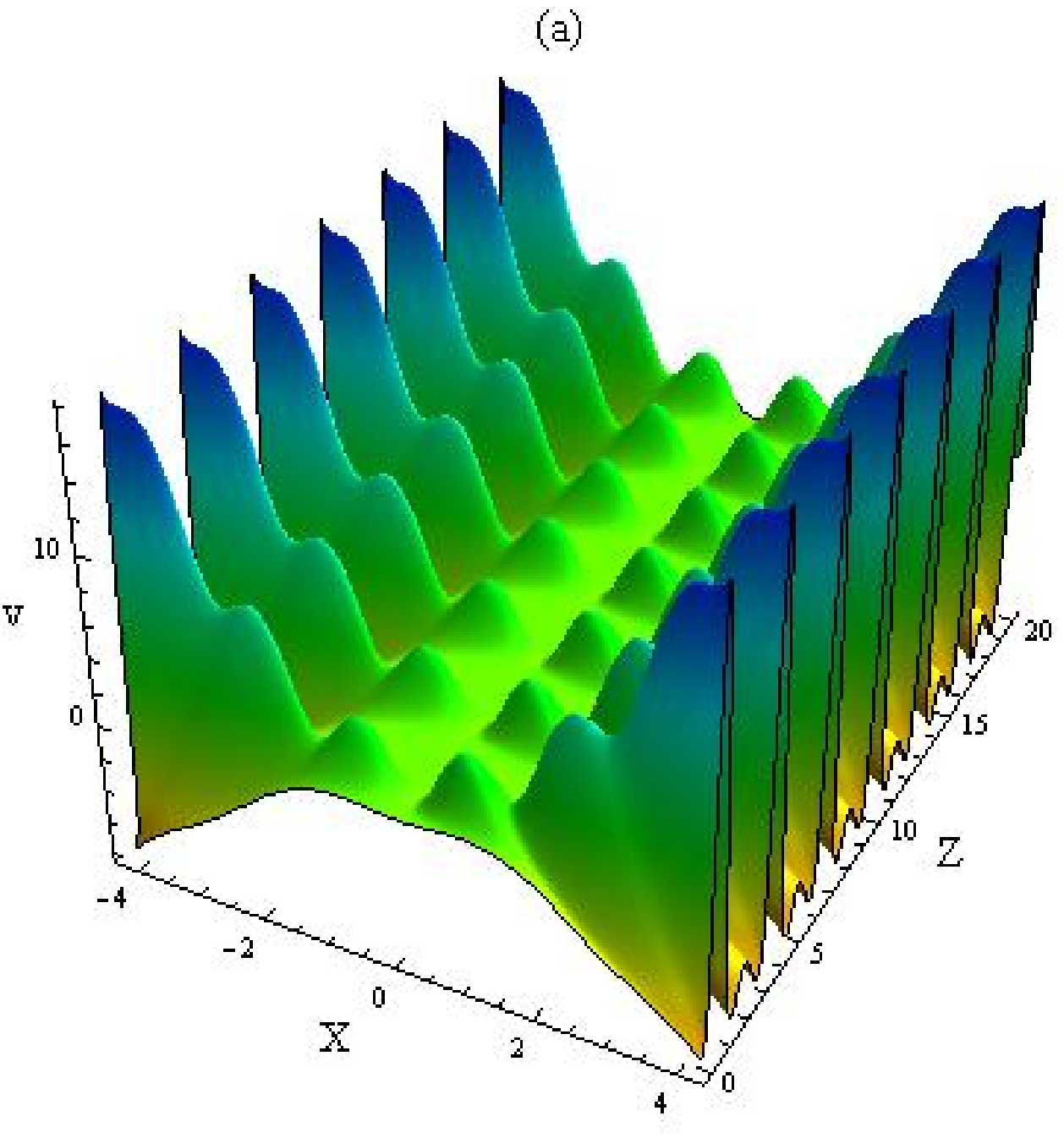}\qquad\qquad %
\includegraphics[width=7.0cm]{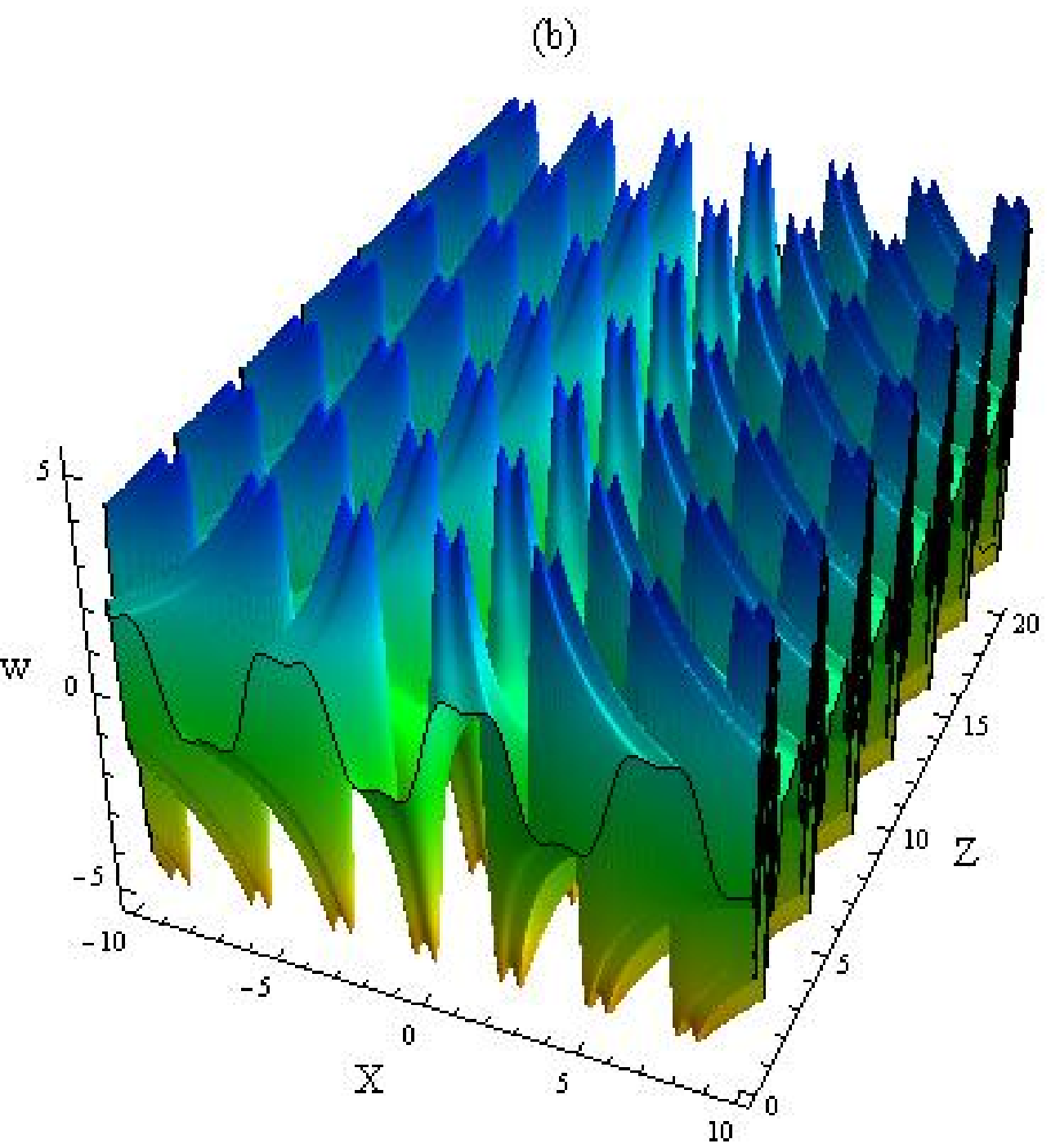}
\caption{Periodic evolution of (a) the even trapping potential $v\left(
X,Z\right) $ and of (b) the odd gain/loss coefficient $w\left( X,Z\right) $,
Eq. (\protect\ref{eq25}) with $\protect\nu =2,$ $\protect\epsilon =2,$ $%
V_{0}=0.5,$ $W_{0}=0.5$, $k=0.4$ and $\protect\gamma _{0}=2.5$. }
\label{fig5}
\end{figure}

In the particular case $h\left[ \xi \right] =1$ the trapping potential $%
v(X,Z)$ and the gain/loss coefficient $w\left( X,Z\right) $ are%
\begin{eqnarray}
v\left( X,Z\right) &=&\frac{\nu ^{2}(\cos \left( 2\nu Z\right) -3-2\epsilon
\cos \left( \nu Z\right) )}{8\left( \epsilon +\cos \left( \nu Z\right)
\right) ^{2}}X^{2}+\frac{\gamma _{0}^{2}}{\left( \epsilon +\cos \left( \nu
Z\right) \right) ^{2}}\times  \notag \\
&&\times \mathrm{sn}^{2}\left( \frac{\gamma _{0}}{\epsilon +\cos \left( \nu
Z\right) }X,k\right) \left( V_{0}+k^{2}W_{0}^{2}\mathrm{sn}^{2}\left( \frac{%
\gamma _{0}}{\epsilon +\cos \left( \nu Z\right) }X,k\right) \right) ,  \notag
\\
w\left( X,Z\right) &=&\frac{\nu \sin \left( \nu Z\right) }{2\left( \epsilon
+\cos \left( \nu Z\right) \right) }+\frac{W_{0}\gamma _{0}^{2}}{\left(
\epsilon +\cos \left( \nu Z\right) \right) ^{2}}\times  \notag \\
&&\times \mathrm{sn}\left( \frac{\gamma _{0}}{\epsilon +\cos \left( \nu
Z\right) }X,k\right) \left( 4\mathrm{dn}^{2}\left( \frac{\gamma _{0}}{%
\epsilon +\cos \left( \nu Z\right) }X,k\right) +k^{2}-1\right) .
\label{eq25}
\end{eqnarray}%
The plots of the $v\left( X,Z\right) $ and $w\left( X,Z\right) $ are shown
in Fig. (\ref{fig5}).

\begin{figure}[h]
\centering
\includegraphics[width=7.0cm]{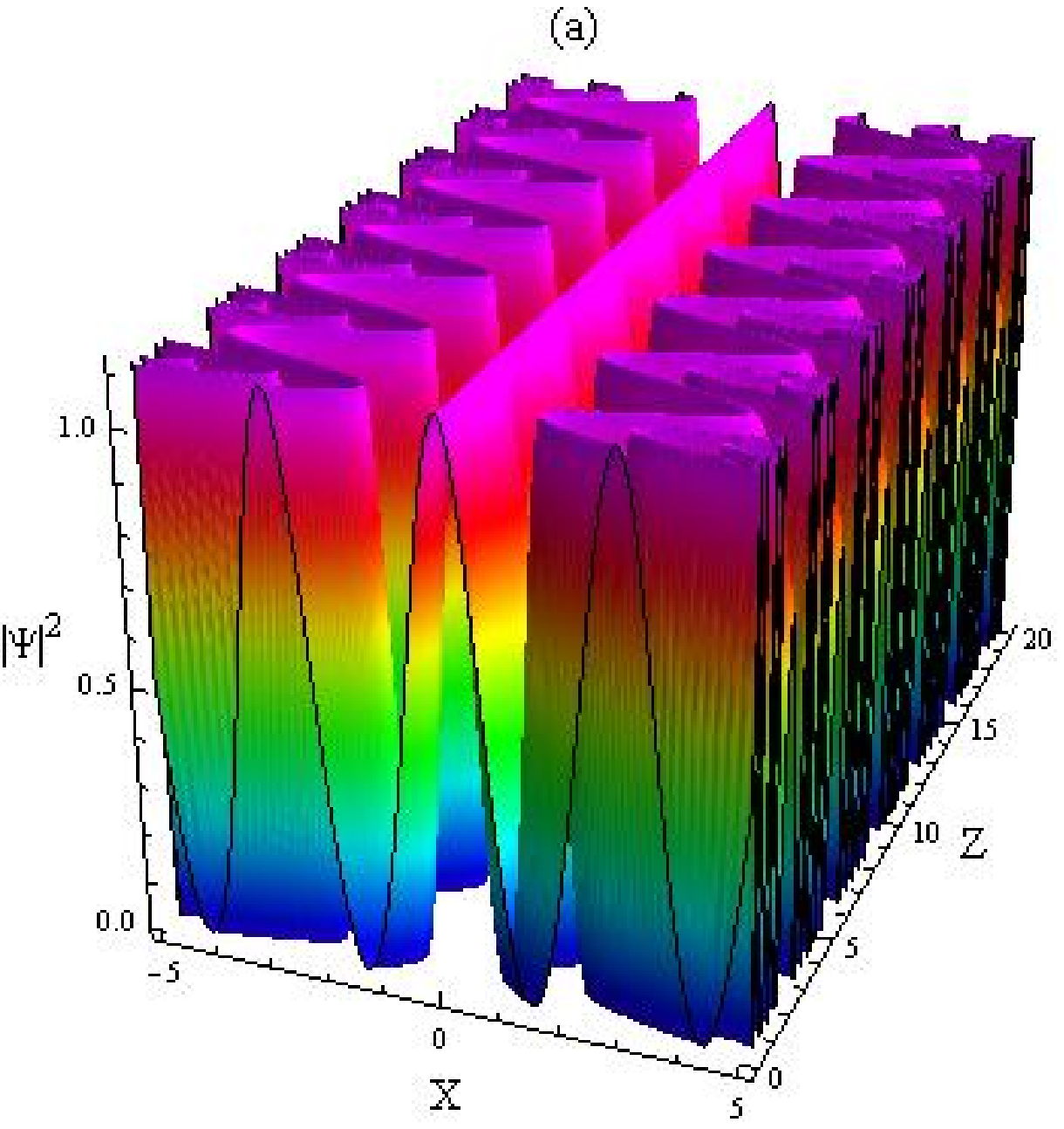}\qquad\qquad %
\includegraphics[width=7.0cm]{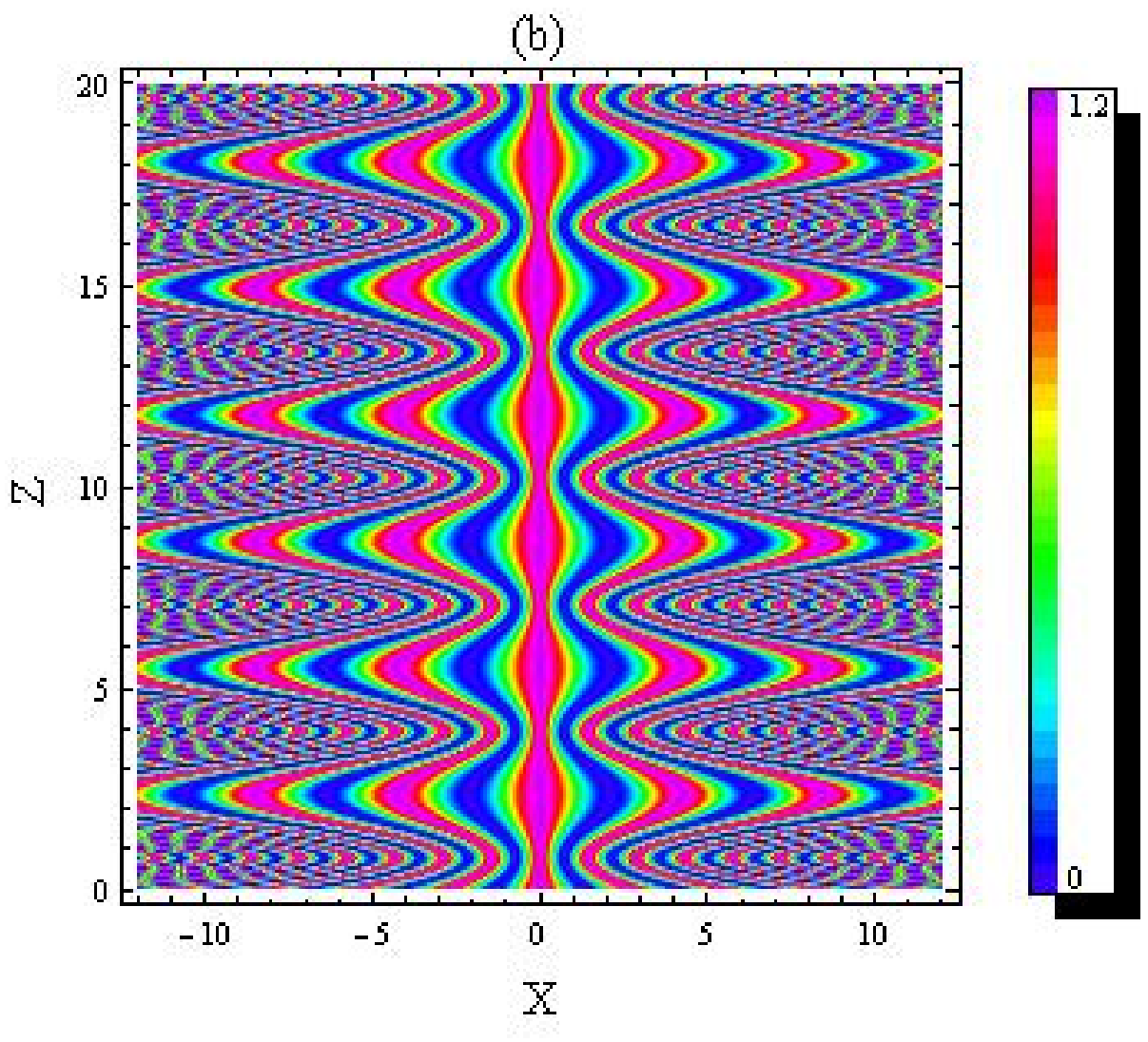}
\caption{Profile (a) and Density plot (b) of $\left\vert \Psi \right\vert
^{2}$, Eq. (\protect\ref{eq26}) with $\protect\nu =2,$ $\protect\epsilon =2,$
$\protect\gamma _{0}=2.5,$ $V_{0}=0.5,$ $W_{0}=0.5$ and $k=0.4$. }
\label{fig6}
\end{figure}

\noindent The wavefunction $\Psi \left( X,Z\right) $, which is solution of
Eq. (\ref{eq16a}) is obtained by substituting Eq. (\ref{eq23}) into the Eq. (%
\ref{eq16b})%
\begin{equation}
\left\vert \Psi \left( X,Z\right) \right\vert =\sqrt{%
V_{0}+W_{0}^{2}k^{2}+W_{0}^{2}+2k^{2}}\left\vert \mathrm{cn}\left( \frac{%
\gamma _{0}}{\epsilon +\cos \left( \nu Z\right) }X,k\right) \right\vert .
\label{eq26}
\end{equation}%
The plot of the $\left\vert \Psi \left( X,Z\right) \right\vert ^{2}$ is
shown in Fig. (\ref{fig6}).

\section{Conclusion}

\label{con} In this paper we have presented a method of obtaining exact
solutions of the NLSE with a cubic inhomogeneous nonlinearity and in the
presence of an external potential. The method has been used to obtain
localized exact solutions CNLSE with inhomogeneous nonlinearity in the
presence of a couple of $\mathcal{PT}$ symmetric external potentials.
However the method is general enough and can be applied to obtain exact
solutions of other potentials if the solution associated to NLSE with
constant cubic nonlinearity is known. We would like to mention that whether
the original CNLSE Eq. (\ref{eq1}) will be $\mathcal{PT}$ symmetric or not
depends on several factors, the crucial among them being the choice of the
function $\gamma (Z)$. Here we have chosen a particular form of $\gamma (Z)$
although there are many other choices which would ensure $\mathcal{PT}$
symmetry of Eq. (\ref{eq1}). In this context it may be noted that besides $%
\mathcal{PT}$ symmetric potentials, there are also $\eta $ pseudo Hermitian
potentials for which the linear Schrödinger Hamiltonian has real eigenvalues
\cite{mostafa}. We feel it would be interesting to try to find exact
solutions of NLSE in the presence of external $\eta $ pseudo Hermitian
potentials. Finally we would like to mention that it will be of interest to
extend the present formalism to higher dimensional \cite{shi} or radially
symmetric NLSE with variable strength management and in the presence of a $%
\mathcal{PT}$ symmetric potential.

\begin{center}
\textbf{Acknowledgement}
\end{center}

LEAM thanks to the CAPES/CNPq-IEL Nacional-Brasil program for the
scholarship. PR wishes to thank UNESP for supporting a visit to UNESP-Campus
de Guaratinguetá during which this work started. This work is also partially
supported by CNPq (procs. 482043/2011-3, 304252/2011-5 and 306316/2012-9).

\end{document}